\documentclass[12pt]{article}
\usepackage{esfconf}
\newcommand{\Purple}[1]{#1}
\newcommand{\Red}[1]{#1}
\newcommand{\Blue}[1]{#1}

\newcommand{\ft}[2]{{\textstyle\frac{#1}{#2}}}
\def\rmi{{\rm i}}
\usepackage{epsf,amssymb,lscape}
\newsavebox{\nwsea}
\sbox{\nwsea}
    {\setlength{\unitlength}{0.825em}
     \begin{picture}(3,3)
        \put(3,0){\vector(-3,2){3}}
        \put(0,2){\vector(3,-2){3}}
        \put(2,1){\mbox{$\textrm{T}_{\rm st}$}}
     \end {picture}}
\newcommand{\nwsearrow}{\mathord{\!\usebox{\nwsea}}}
\newsavebox{\neswa}
\sbox{\neswa}
    {\setlength{\unitlength}{0.825em}
     \begin{picture}(3,3)
        \put(0,0){\vector(3,2){3}}
        \put(3,2){\vector(-3,-2){3}}
        \put(2.6,0.6){\mbox{$\textrm{T}_{\rm st}$ }}
     \end {picture}}
\newcommand{\neswarrow}{\mathord{\!\usebox{\neswa}}}
\newsavebox{\nsa}
\sbox{\nsa}
    {\setlength{\unitlength}{0.825em}
     \begin{picture}(3,3)
        \put(1.9,0){\vector(0,1){2}}
        \put(1.9,2){\vector(0,-1){2}}
        \put(2.4,1){\mbox{$\textrm{T}_{\rm tt}$}}
        \put(0,1){\mbox{$\textrm{T}_{\rm ss}$}}
     \end {picture}}
\newcommand{\nsarrow}{\mathord{\!\usebox{\nsa}}}

\begin{document}
\begin{titlepage}
\begin{flushright}
KUL-TF-2000/22\\ UG-00-13\\
hep-th/0010194
\end{flushright}
\vspace{.5cm}
\begin{center}
\baselineskip=16pt
{\LARGE   The unifying superalgebra $OSp(1|32)$  
}\\
\vfill
{\large Eric Bergshoeff$^1$ and Antoine Van Proeyen $^{2,\dagger }$, 
  } \\
\vfill
{\small
$^1$  Institute for Theoretical Physics, Nijenborgh 4,
9747 AG Groningen, The Netherlands,
\\ \vspace{6pt}
$^2$ Instituut voor
Theoretische Fysica, Katholieke Universiteit Leuven, Celestijnenlaan
200D, B-3001 Leuven, Belgium.
}
\end{center}
\vfill
\begin{center}
{\bf Abstract}
\end{center}
{\small
We show how $OSp(1|32)$ gives a unifying framework to describe $d=10$
type~II string theories, $d=11$ M-theory and $d=12$ F-theory. The
theories are related by different identifications of their symmetry
operators as generators of $OSp(1|32)$. T- and S-dualities are recognized
as redefinitions of generators. Some $(s,t)$ signatures of spacetime
allow reality conditions on the generators. All those that allow a real
structure are related again by redefinitions within the algebra, due to
the fact that the algebra $OSp(1|32)$ has only one real realization. The
redefinitions include space/space, time/time and space/time dualities. A
further distinction between the theories is made by the identification of
the translation generator. This distinguishes various versions of type~II
string theories, in particular the so-called $*$-theories, characterized
by the fact that the $P_0$ generator is not the (unique)
positive-definite energy operator in the algebra.
\vspace{2mm} \vfill \hrule width 3.cm}
To be published in the proceedings of the International Conference \textit{`Quantization,
Gauge Theory and Strings'} dedicated to the memory of Professor Efim
Fradkin, Moscow, June 2000.\\[5mm]
{\footnotesize
\noindent $^\dagger$ Onderzoeksdirecteur FWO, Belgium }
\end{titlepage}


\title{The unifying superalgebra $OSp(1|32)$ }


\authors{E. Bergshoeff\adref{1} and A. Van Proeyen\adref{2}}

\addresses{\1ad Institute for Theoretical Physics, Nijenborgh 4,
9747 AG Groningen, The Netherlands, \nextaddress \2ad Instituut voor
Theoretische Fysica, Katholieke Universiteit Leuven, Celestijnenlaan
200D, B-3001 Leuven, Belgium.}

\maketitle

\begin{abstract}
We show how $OSp(1|32)$ gives a unifying framework to describe $d=10$
type~II string theories, $d=11$ M-theory and $d=12$ F-theory. The
theories are related by different identifications of their symmetry
operators as generators of $OSp(1|32)$. T- and S-dualities are recognized
as redefinitions of generators. Some $(s,t)$ signatures of spacetime
allow reality conditions on the generators. All those that allow a real
structure are related again by redefinitions within the algebra, due to
the fact that the algebra $OSp(1|32)$ has only one real realization. The
redefinitions include space/space, time/time and space/time dualities. A
further distinction between the theories is made by the identification of
the translation generator. This distinguishes various versions of type~II
string theories, in particular the so-called $*$-theories, characterized
by the fact that the $P_0$ generator is not the (unique)
positive-definite energy operator in the algebra.
\end{abstract}


\section{Introduction}

The group $OSp(1|32)$ was already mentioned in the first papers on $d=11$
supergravity~\cite{CremmerJulia}. This algebra and its extension
$OSp(1|64)$ appeared as anti-de Sitter (adS) and superconformal algebras
in $d=10$ and $d=11$ Minkowski theories~\cite{vanHolten:1982mx} long ago,
and got new attention related to the $M$-theory algebra~\cite{Malgebra}.
The adS or conformal algebras got new attention in a recent paper on the
superconformal aspects of $d=11$ theories \cite{West:2000ga} and in
two-time theories \cite{BDM2t}. In these two cases, the
$OSp(1|64)$ conformal group appeared. In the physical theories that we
consider, we need the subgroup of $OSp(1|64)$ that is a contraction of
$OSp(1|32)$ in a way that will be clarified below.

Our initial motivation to study the role of the $OSp(1|32)$ algebra was
related to Euclidean theories. When one considers the $D$-instanton
\cite{Dinstanton}, one often considers the bosonic theory, ignoring its
possible embedding in the supersymmetric theory. In particular, one makes
use of the IIB theory in Euclidean space, while the latter can not be
formulated as a supersymmetric theory with real fields, as we will show
below. Remark that the connection between these Euclidean theories and
the Minkowski string theories involve a duality between theories of
different spacetime signature \cite{dualst}.

A second question that was posed when we started this research, was
related to the observation that in many super-Euclidean theories one makes
use of complexification of the fields and in other cases one does not
\cite{Euclidean}. We would like to know when it is necessary to do so, and
when it can be avoided.

Apart from the possibility of no time directions, one is also interested
in theories with more time directions \cite{Hull,BDM2t,StarTheories}.
Therefore, it looked natural to extend our investigation to an arbitrary
spacetime signature.

This leads to a web of dualities between theories in $d=10$, 11 and 12 of
different spacetime signature, similar to what has been found in
\cite{Hull}. We obtain these dualities from an algebraic approach, which
puts the contraction of $OSp(1|32)$ as a unifying principle. The different
theories are then just many faces of the same underlying symmetry group.
This seminar summarizes the results obtained in \cite{faces}. In
\cite{GurseyFaces} we clarify the relation between the super-Poincar\'{e}
algebra that we consider here, and the full $OSp(1|32)$ as super-adS
algebra or $OSp(1|64)$ as superconformal algebra.
\par
Our results are divided in 3 main parts. In
section~\ref{ss:complexalgebras} we consider the complex algebra and its
realizations in the different dimensions, in
section~\ref{ss:realalgebras} we consider the real algebra and its
realizations in different spacetime signatures, and in
section~\ref{ss:TranslationEnergy} we identify the translation operator,
distinguishing between different Lagrangian theories for the same
spacetime signature. Throughout the work we indicate the dualities
connecting all the theories. Finally, a short summary is given in
section~\ref{ss:conclusions}.

\section{Complex symmetry algebras}\label{ss:complexalgebras}

$OSp(1|32)$ is the algebra of 32 fermionic charges with all possible
bosonic generators in their anticommutator. We recapitulate its
definition, with the bosonic generators defining $Sp(32)$. Then we will
see how the contraction, explained in more detail in \cite{GurseyFaces},
underlies the F-theory of 12 dimensions, the M-theory of 11 dimensions,
and the IIA and IIB string theories in 10 dimensions. They are obtained
by identifying appropriate subgroups of $Sp(32)$ as the Lorentz
rotations. Note that in the case of the extended algebras mentioned in
\cite{GurseyFaces}, this $Sp(32)$ is the automorphism algebra of the
supersymmetries, in the semi-direct product with $OSp(1|32)$. In any
case, the supersymmetries should be in a spinor representation of the
Lorentz group. Dimensional reduction and T-dualities are then obtained as
mappings between generators of $OSp(1|32)$.

The algebra $OSp(1|32)$ is given by
\begin{eqnarray}\label{commMQ}
&&\left\{ \Red{Q_A},\Red{Q_B}\right\}=\Blue{Z_{AB}}\,,\qquad
  \left[\Blue{Z_{AB}},\Red{Q_C}\right]= \Red{Q_{(A}}\Omega_{B)C}\,,\nonumber\\
&&  \left[ \Blue{Z_{AB}},\Blue{Z_{CD}}\right] =
\Omega_{A(C}\Blue{Z_{D)B}}+ \Omega_{B(C}\Blue{Z_{D)A}}\,,
\end{eqnarray}
where $A$ runs over 32 values, $Z_{AB}$ is symmetric and has thus $\ft12 .
32.33= 528$ components. $\Omega _{AB}$ is an antisymmetric invertible
metric, and as such, the last commutator defines $Sp(32)$.

To recognize this algebra as a symmetry algebra in $d$ dimensions, one has
to embed $SO(d)$ in $Sp(32)$, in such a way that the spinor
representation of $SO(d)$ fits in the 32. This makes already clear that
$d=12$ is the highest possible dimension. To make that identification, we
have to select chiral spinors $\Red{\hat{Q}}$ of 12 dimensions. These are
defined using the chiral projection $\hat{{\cal P}}^+$:
\begin{equation}
\hat{{\cal P}}^+  \Red{\hat{Q}}= \Red{\hat{Q}}\,,\qquad \hat{{\cal
P}}^+=\ft12(1+\hat{\Gamma} _*)\,,\qquad \hat{\Gamma} _*=\Gamma _1\Gamma
_2\ldots \Gamma_{12}\,.
 \label{chirald12}
\end{equation}
Remark that we use the notation $\Gamma _*$ (the hat specifies the
12-dimensional context) in any even dimension to denote the product of all
the gamma matrices, similar to $\gamma _5$ in 4 dimensions. Then the
algebra (\ref{commMQ}) is realized by identifying $\Omega _{AB}$ with
${\cal C}_{\alpha \beta }$, the (antisymmetric) charge conjugation matrix
of $d=12$. Splitting the matrix $Z_{AB}$ in its irreducible
representations, the anticommutator of the supersymmetries looks like
\begin{equation}
\begin{array}{ccccc}
   \left\{\Red{\hat{Q}},\Red{\hat{Q}} \right\}& =& \ft12 \hat{{\cal P}}^+ \hat{\Gamma
}^{\hat{M}\hat{N}}\Blue{\hat{Z}_{\hat{M}\hat{N}}} & +&\ft1{6!} \hat{{\cal
P}}^+ \hat{\Gamma }^{\hat{M}_1\cdots
\hat{M}_6}\Blue{\hat{Z}^+_{\hat{M}_1\cdots \hat{M}_6}}\,,
\\[2mm]
  \frac12 .32.33 & = &\ft12 .12.11& + & \frac12\,\,
  \frac{12.11.10.9.8.7}{1.2.3.4.5.6}\,,
\end{array}\label{algd12}
\end{equation}
where the last line shows that indeed all 528 generators are present, and
the right-hand side thus contains everything which is consistent from the
symmetry property of an anticommutator.

In 11 dimensions, the bosonic generators split as $528=11+55+462$,
following the anticommutator
\begin{equation}
   \left\{ Q_\alpha,\,Q_\beta\right\}=
  \Gamma^\mu _ {\alpha\beta} P_\mu   +
 2\Gamma^{\mu\nu}_{\,\alpha\beta} Z_{\mu\nu}
+\frac{1}{5!}\Gamma^{\mu\nu\rho\sigma\tau}_{\,\alpha\beta}
Z^5_{\mu\nu\rho\sigma\tau}\,.
 \label{QQd11}
\end{equation}
In 10 dimensions one can
again define chiral spinors, which are 16-dimensional, and consider either
2 generators of opposite chirality (IIA) or of the same chirality (IIB).
In the first case, the anticommutators are
\begin{eqnarray}
\left\{Q^{\pm },Q^{\pm } \right\} & = & {\cal P}^{\pm }\Gamma ^M Z^{\pm
}_M+\ft 1{5!}{\cal P}^{\pm }\Gamma ^{M_1\cdots M_5}Z^\pm _{M_1\cdots
M_5}\, ,
\nonumber\\
\left\{Q^{\pm },Q^{\mp } \right\} & = & \pm {\cal P}^{\pm } Z +
\ft12{\cal P}^{\pm }\Gamma ^{MN}Z_{MN} \pm \ft 1{4!}{\cal P}^{\pm }\Gamma
^{M_1\cdots M_4}Z _{M_1\cdots M_4}\,. \label{calgIIA}
\end{eqnarray}
The 528 generators are thus split as $2\times (10+126)$ in the
anticommutators between generators of the same chirality and $1+45+210$
in the anticommutator between generators of opposite chirality.

For the IIB case, we have a doublet of fermionic generators $Q^i$,
$(i=1,2)$, of the same chirality, and the anticommutators are
\begin{eqnarray}
  \left\{Q^i,Q^j \right\}&=&{\cal P}^+\Gamma ^MY_M^{ij}
     +\ft1{3!}{\cal P}^+\Gamma ^{MNP}\varepsilon ^{ij}Y_{MNP}
  +{\cal P}^+\ft1{5!}\Gamma ^{M_1\cdots M_5 }Y_{M_1\cdots M_5 }^{+\,ij}\,,
\nonumber\\
Y_M^{ij} & = &  \delta ^{ij}Y_M^{(0)}
   +\tau _1^{ij}   Y_M^{(1)}+\tau _3^{ij}   Y_M^{(3)}\,,
\label{calgIIB}
\end{eqnarray}
where in the second line we have split the symmetric matrix $Y^{ij}$ in
three components, as we can also do for the 5-index generators. The
decomposition is here $528=(3\times 10)+120+(3\times 126)$.

It is clear that all these algebras are related. The dimensional
reductions relate the generators as follows. The chiral generator
$\hat{Q}$ of 12 dimensions, splits in 10 dimensions in a chiral and an
antichiral generator, as follows from the relation $\hat{\Gamma
}_*=\Gamma _*\otimes \sigma _3$ for a convenient realization of gamma
matrices, where $\Gamma _*$ is the product of 10 gamma matrices of
dimension $32\times 32$ in 10 dimensions (for the realization that we use
in any dimension see \cite{tools}). The two chiral generators are the
$Q^\pm $ in (\ref{calgIIA}), and adding them gives the 32-component
generator $Q=Q^++Q^-$ used for $d=11$.  The T-dual theories are
identified by taking
\begin{equation}
  Q^+= Q^1\,,\qquad Q^-=\Gamma^sQ^2\,,
 \label{QTdual}
\end{equation}
where $\Gamma ^s$ is a gamma matrix in an arbitrary (spacelike or
timelike) direction. On the other hand, S-duality is the mapping
\begin{equation}
  Q^i  \stackrel{S}{\longrightarrow}  \left( e^{\rmi\ft14 \pi \tau _2}\right)
{}^i{}_jQ^j\,.
 \label{QSdual}
\end{equation}
Thus all the dimensional reductions and dualities are written as mappings
between the generators of $OSp(1|32)$. We  mentioned here only the
fermionic generators explicitly, as the rules for the bosonic generators
follow from identifying the anticommutator relations before and after the
map. E.g. when going from 12 to 11 dimensions, this leads to the
identifications
\begin{equation}
\tilde{Z}_{\tilde{M}}  =  \rmi \hat{Z}_{\tilde{M}\,12}\, , \qquad
\tilde{Z}_{\tilde{M}\tilde{N}}  =  \hat{Z}_{\tilde{M}\tilde{N}}\, ,\qquad
\tilde{Z}_{\tilde{M}_1\cdots \tilde{M}_5}=2\rmi
\hat{Z}_{\tilde{M}_1\cdots \tilde{M}_5\,12}\, . \label{ident1211}
\end{equation}
Note that the appearance of factors $\rmi$ is irrelevant here, as we can
make redefinitions of generators involving $\rmi$ at random. For the real
algebras, to be discussed below, this should be possible consistently
with the reality conditions, as we checked in \cite{faces}.

\section{Real symmetry algebras}\label{ss:realalgebras}

The important fact for the real forms is the uniqueness of the real form
of the superalgebra $OSp(1|32)$. Therefore the equivalences of all the
symmetry algebras of section~\ref{ss:complexalgebras} are valid also for
the real form, when it exists. The real form  exists only for specific
spacetime signatures. The dimensional reduction and T-duality acts now
between theories of specific signatures. We have to distinguish then
space/space, time/time and space/time dualities.

Considering the table of real forms of the basic Lie superalgebras (see
e.g.  table~5 of \cite{tools} for a convenient presentation), we see that
nearly all superalgebras have different real forms, even the exceptional
superalgebras. But the algebras $OSp(1|n)$ have only one real form, with
$Sp(n,\mathbb{R})$ as bosonic subalgebra.

To realize this real algebra in $d$ dimensions, we have to consider when
we can impose consistent reality conditions on the fermionic generators.
This is sufficient to be able to classify all the realizations of the
unique real superalgebra $OSp(1|32)$. Table~2 of \cite{tools} gives the
summary of the results that we need. We need 32 real supercharges. The
table shows that $d=12$ with $(space,time)$ signature $(10,2)$ is the
highest possible dimension. In general the results are invariant under
$(s,t)\simeq (s-4,t+4)$, thus $(6,6)$ is also possible. The interchange
of $s$ and $t$ is irrelevant, and corresponds merely to a change of
notations of mostly $+$ to mostly $-$ metrics. Therefore we do not
mention the $(2,10)$ signature. To make the projections to real spinors
one uses three types of projections, Weyl, Majorana or symplectic
Majorana:
\begin{eqnarray}
 W & : & Q=\Gamma _*Q \nonumber\\
 M & : & Q^*= \alpha B Q \,,\quad B\equiv -{\cal C}\Gamma _1\ldots \Gamma _t\,, \nonumber\\
 &&BB^*=1\,,\quad\alpha \alpha ^*=1\,, \nonumber\\
 SM&:&BB^*=-1\,,\quad\alpha \alpha ^*=-1\,,\qquad\alpha \mbox{ antisymmetric matrix.}
 \label{MWS}
\end{eqnarray}
The first one is the Weyl projection to chiral spinors that we already
encountered in (\ref{chirald12}). The other two are reality conditions.
$B$ is a definite matrix in spinor space, which is defined here using the
charge conjugation matrix ${\cal C}$, and all the timelike $\Gamma
$-matrices.
 On the other hand, $\alpha $ is a
scalar in the spinor space, but may be a matrix acting on the different
generators $Q$, e.g. between $Q^+$ and $Q^-$ in type~IIA, or between
$Q^1$ and $Q^2$ in type~IIB. In the cases where the table indicates `M',
$BB^*=1$ and a consistent reality condition can be obtained with $\alpha
\alpha ^*=1$. Therefore in these cases, a single spinor can suffice, and
these are the so-called Majorana spinors. In other cases $BB^*=-1$,
consistency (taking the $*$ of a $*$ is an identity) also $\alpha \alpha
^*=-1$. Therefore in these cases we need a matrix $\alpha $ of even
dimension. We thus have a doubling of the generators, and this is the
so-called symplectic Majorana condition. This leads to the possibilities
for 32-component real spinors, listed in table~\ref{tbl:MWSspinors}.
\begin{table}[ht]
\tabcolsep 2pt
\begin{center}
\begin{tabular}{|c|cccccccccccc|}
\hline
12  & &&(10,2) &  &  &  & &&&& (6,6) &  \\
64  & &&MW     &  &  &  & &&&&MW &  \\[3mm]
11  & &(10,1) & & (9,2) &  & \phantom{(9,2)} &&\phantom{(9,2)}&& (6,5)&=&(5,6)   \\
32  & &M      & & M     &  &  &&&&  M &&M  \\[3mm]
10  & (10,0) && (9,1)  && (8,2)&& (7,3) && (6,4) && (5,5)& \\
32  & SM     && MW     && M    && SMW   &&  SM   && MW &\\
    & A      && A/B    && A    &&  B    &&  A    && A/B&\\
\hline
\end{tabular}\tabcolsep 6pt
  \caption{\sl The possible spacetime signatures for 32 real spinor generators.
The first column indicates the number of complex generators that are
present before any projection. The last row indicates, for each
signature, whether in $d=10$ a real form for type~IIA (A), type~IIB (B)
or both (A/B) exists.
 }\label{tbl:MWSspinors}
\end{center}
\end{table}

\begin{figure}[ht]
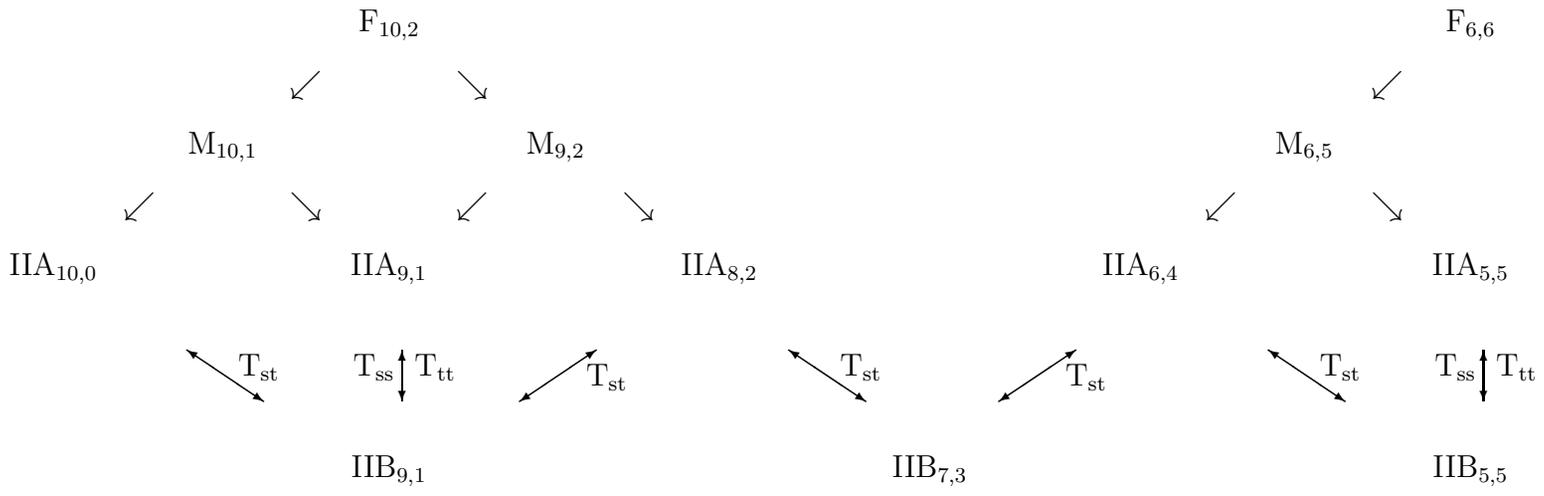

\[
\begin{array}{cccccccccccccccccccc}
  &  &  &          &      \textrm{F}_{10,2}          &           &  &  &  &  &  &  &  &  &  &  &   \textrm{F}_{6,6} \\[3mm]
  &  &  &           \swarrow &   &\searrow              &  &  &  &  &  &  &  &  &  &   \swarrow  &  \\[3mm]
            &     &  \textrm{M}_{10,1} &                  &   &       & \textrm{M}_{9,2} &     &&  &  &  &  &  &    \textrm{M}_{6,5}&  &   \\[3mm]
            &\swarrow &     &\searrow&                 &  \swarrow&       &\searrow &    &  &  &  &  & \swarrow &    & \searrow  &    &  &  &  \\[3mm]
\textrm{IIA}_{10,0}  &   &  &  & \textrm{IIA}_{9,1}&&  &  & \textrm{IIA}_{8,2} &  &  &  & \textrm{IIA}_{6,4} &  &  &  & \textrm{IIA}_{5,5} &  \\[5mm]
            &   &  \nwsearrow     &  & \nsarrow  &&   \neswarrow      & &  &  \nwsearrow       &  &  \neswarrow &  &  & \nwsearrow   &  & \nsarrow  \\[5mm]
            &   &                 &  & \textrm{IIB}_{9,1} &  &  &  &  &  & \textrm{IIB}_{7,3} &    &  &  & & & \textrm{IIB}_{5,5}    \\[3mm]
\end{array}
\]
\caption{The dimensional reductions and T-dualities between the real
algebras  in $d=12$, 11 and 10 of different signatures. The diagonal
one-sided arrows indicate dimensional reductions. The  both-sided arrows
refer to the space/space ($\rm T_{\rm ss}$), time/time ($\rm T_{\rm tt}$)
and space/time ($\rm T_{\rm st}$) dualities discussed in the text.
\label{ff:small}   }
\end{figure}

One can then consider the dimensional reductions and T-dualities
discussed in section~\ref{ss:complexalgebras}, but now we have to be
careful with the signatures. When performing the T-duality as in
(\ref{QTdual}), one has to distinguish whether $\Gamma ^s$ is a timelike
or a spacelike gamma matrix. This $s$-direction can even be timelike for
the IIA algebra and spacelike for the IIB algebra or vice--versa. These
are the time/space or space/time T-dualities, changing the signature.
This leads to the both-sided arrows in figure~\ref{ff:small}. In
\cite{faces} it was shown that all these identifications of algebras
require mappings between the $\alpha $ factors in (\ref{MWS}) for the
different realizations. At the end, these are all redefinitions of the
generators. The reality conditions on the fermionic generators lead to
reality conditions on the bosonic generators in order that the
anticommutation relations are consistent. The dimensional reduction and
T-dualities give mappings that relate real bosonic generators in one
algebra with real bosonic generators in the other algebra. This is highly
nontrivial, but it is bound to work out, due to the uniqueness of the
real form of $OSp(1|32)$.

\landscape\tabcolsep 1pt
\tabcolsep 6pt

\endlandscape

\section{Translations and the energy operator}\label{ss:TranslationEnergy}

In the third step we identify one of the vector generators as
`translations'. This identification is essential for a spacetime
interpretation of the theory. The different possibilities for this
identification distinguish e.g. IIA from IIA$^*$ theories. We will then
remark that T-duality gives a mapping between different types of
generators. It can mix e.g. translations with `central charges'. Finally,
we will see that there is a unique positive energy operator in the
algebra. However, that generator is not always the timelike component of
the translations. For instance, in IIA theories, the positive operator is
$P_0$, but in IIA$^*$ theories it is another one, and thus $P_0$ is not
positive in that case.

So far, all bosonic generators were treated on equal footing. To make the
connection between algebras and a spacetime theory, we want to know which
generator performs `translations' in spacetime. Seen in another way,
spacetime is the manifold defined from a base point by the action of this
`translation' generator. This is thus similar to the coset space idea. To
generate a spacetime of the appropriate dimension, the translation
operator should be a vector operator in the theory. This is nearly the
only requirement, apart from a non-degeneracy condition. Indeed, in order
that the supersymmetries perform their usual role, they should square to
the translations. Thus the matrix that appears in the anticommutator
between all the supersymmetries, defining how they square to
translations, should be non-degenerate.

For $d=12$, with the algebra (\ref{algd12}), there is no vector operator.
Thus there is no candidate for translations, implying that F-theory has
no straightforward spacetime interpretation. On the other hand, for
$d=11$, with the algebra (\ref{QQd11}), there is one vector operator, and
this one should thus be called the translation generator.

In 10 dimensions it becomes more interesting. Consider first the IIA
algebra (\ref{calgIIA}). There are 2 vector operators $Z_M^+$ and
$Z_M^-$. Both separately are not convenient, because then one half of the
supersymmetries would not square to translations. But we can take linear
combinations. For the signature $(9,1)$ there are, up to redefinitions,
two choices consistent with the reality conditions
\begin{eqnarray}
(9,1)\ :\ {\rm IIA}\phantom{*} & : & \Purple{P_M}\equiv \Red{Z_M^+} + \Red{Z_M^-}\,, \nonumber\\
{\rm IIA}^* & : &\Purple{P_M}\equiv \Red{Z_M^+} - \Red{Z_M^-}\,.
\label{defIIA*}
\end{eqnarray}
We label these choices as IIA and IIA$^*$ in accordance with \cite{Hull}.
For signatures $(10,2)$ or $(8,2)$ there are the possibilities
\begin{eqnarray}
(10,0)\mbox{ or }(8,2)\ :\ \textrm{IIA}& : &
\Purple{P_M}\equiv \rmi(\Red{Z_M^+} + \Red{Z_M^-})\,,\nonumber\\
  \textrm{IIA}'& :& \Purple{P_M}\equiv \Red{Z_M^+} -   \Red{Z_M^-}\,.
\label{defPIIAp}
\end{eqnarray}
However, now these two choices can be related by a redefinition $Q^\pm
\rightarrow e^{\pm \rmi\pi /2} Q^\pm$. Such a redefinition is, similar to
(\ref{QSdual}) and therefore we recognize it as an S-duality.

The operators that are not translations, remain as `central charges' in
the theory. Therefore we see that the generator that is a translation in
one theory, appears as a central charge in the other theory, as we
announced in the beginning of this section.

For the IIB algebra (\ref{calgIIB}), there are three candidates for
translations. For signature (9,1) they are all consistent with the
reality condition. We thus distinguish
\begin{eqnarray}
(9,1)\ :\hskip .2truecm  && {\rm IIB}\phantom{*} \ : \
\Purple{P_M}=\Red{Y^{(0)}_M}\, , \nonumber\\
&&
{\rm IIB}^* \  : \ \Purple{P_M}=\Red{Y^{(3)}_M}\,, \nonumber\\
&&{\rm IIB}' \  : \ \Purple{P_M}=\Red{Y^{(1)}_M}\,.
 \label{PIIB}
\end{eqnarray}
Considering the possible redefinitions, we come back to (\ref{QSdual}).
This S-duality leaves the IIA translation generator invariant, and
relates the translation of IIB$^*$ with that of ${\rm IIB}'$. On the other
hand, for the signature (7,3) there are three S-dual versions.

\begin{figure}
\begin{center}
\leavevmode \epsfxsize=12cm
 \epsfbox{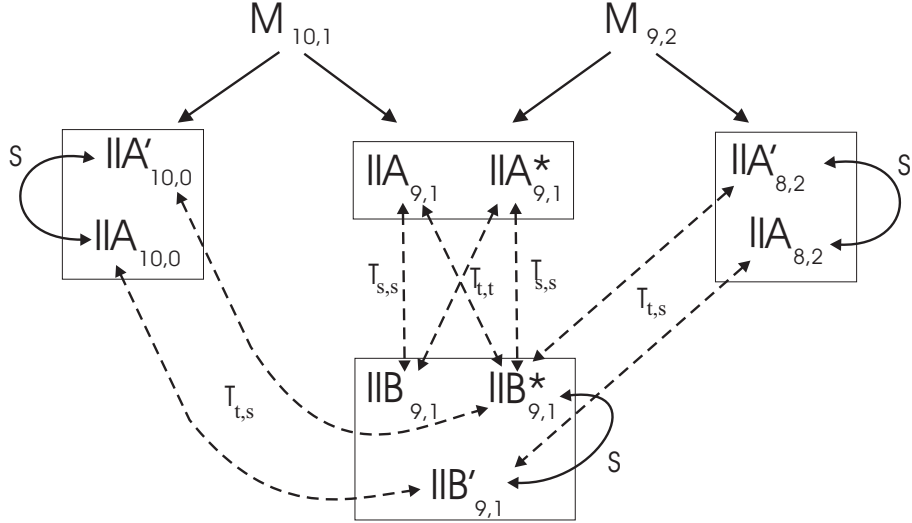}
 \caption{\it Dimensional reduction, T and S-dualities after specification
of the translation operator.\label{fig:detailDRdual}}
\end{center}
\end{figure}

We present the results schematically in figure~\ref{fig:detailDRdual}. It
is a detail of a part of figure~\ref{ff:small}. The specification of the
translation generator distinguishes different IIA and IIB theories. The
T-dualities connect specific versions. S-dual theories are mapped by
T-dualities to S-dual theories.

Finally, we can identify one bosonic operator in $OSp(1|32)$ that is
positive. We can identify this one from the anticommutation relation
\begin{equation}
  \left\{Q^i, Q^{j\,*}\right\} =
  \left\{Q^i, Q^{j}\right\}  B^T \geq
   0\,,
 \label{PositiveQQ}
\end{equation}
using the Majorana condition\footnote{We take here $\alpha =1$ and, to
argue for the positivity, we use the convention that complex conjugation
does not change the order of the operators. See \cite{faces} for the
arguments independent of these conventions} in (\ref{MWS}). We denote
\begin{equation}
  \left\{Q^i, Q^{j}\right\}=  \mathcal{Z}^{ij} {\cal C}^{-1}
 \label{QQZC}
\end{equation}
to represent all the anticommutation relations, where $\mathcal{Z}^{ij}$
is a matrix in spinor space as well as in the $i,j$ indices. For
convenience we write here the charge conjugation matrix explicit. Using
the expression of $B$ in (\ref{MWS}), this implies
\begin{equation}
  \mathcal{Z}^{ij}\Gamma ^t \cdots \Gamma ^1 \geq 0\,.
 \label{Zpos}
\end{equation}
Therefore, the trace of that operator has positive eigenvalues. When we
split $Z$ as usual in different irreducible representations for the
spacetime Lorentz group, then, in order to absorb the gamma matrices, the
relevant part of $Z$ has as many spacetime indices as there are time
directions. All its directions should be timelike. Thus for Minkowski
spaces it is the timelike component of a vector operator, while for
Euclidean theories this positive operator is a scalar `central charge'.
For Minkowski theories we can thus wonder whether the positive energy
operator is the timelike component of the operator that we selected as
`translations'. If this is the case then the usual Hamiltonian will be
positive. When the positive energy is the timelike component of another
vector operator, then the Hamiltonian built from $P_0$ is not positive
definite. This is what happens in the IIA$^*$, IIB$^*$ and IIB$^\prime$
theories. In these theories, the kinetic energies of some of the $p$-form
gauge fields are negative definite. As an example consider the vector
operators in the IIB-like theories, as in (\ref{calgIIB}). With our
convention that $\alpha =1$, the trace in (\ref{Zpos}) selects the $M=0$
component of $\Red{Y_M^{(0)}}$ as the positive definite energy. Thus it
is indeed the IIB theory where the energy is the timelike part of
translations, and not for the other versions. The algebraic approach thus
gives an understanding of the positivity in type IIA and IIB versus lack
of positivity in the other theories.

\section{Conclusions}\label{ss:conclusions}

The algebras of F-theory, M-theory, type IIA and IIB, ... are different
faces of the same superalgebra $OSp(1|32)$. The uniqueness of the real
form of that algebra implies that all these manifestations can be related
by mappings between the  generators of the algebra. That holds especially
for the dimensional reductions, T- and S-dualities that relate these
theories. Different spacetime signatures are easily incorporated.
However, for certain spacetime signatures, some theories may exist only
in complex form. That answers the questions about why we need sometimes a
complexification procedure to obtain an Euclidean theory. In particular,
the IIB theory has no real form in (10,0). Therefore, in order to discuss
the D-instanton in IIB, we have to give up the concept of a theory with
real fields and action.

We have understood the *-theories as being distinguished from the usual
IIA and IIB by a different identification of the translation generator.
They are related to the ordinary theories by a `duality' interchanging
translations with central charges. Thus in these dualities the concept of
spacetime is very intriguing. It should be interchanged with a sort of
harmonic space where coordinates are associated also with other (vector)
central charges. The unique positive energy operator is the timelike
component of the translations in the ordinary type IIA and IIB theories,
but in other versions (*-theories or theories with a different
signature), it is not the $P_0$ operator that is positive, but rather a
component of a central charge operator.

\noindent{\bf Acknowledgements.}

We have enjoyed a very stimulating atmosphere at the Memorial conference
for
Prof. Fradkin and thank the organizers for providing this opportunity.
This work was supported by the European Commission TMR programme
ERBFMRX-CT96-0045, in which E.B. is associated with Utrecht University.

\end{document}